\documentstyle[epsfig,amssymb,preprint,aps]{revtex}

\begin{document}
\draft
\title{Is Electron an Anyon with Spin-1/2?}
\author{M. Saglam}
\address{Department of Physics, Faculty of Sciences, Ankara University, 06100\\
Tandogan, Ankara. Turkey}
\maketitle

\begin{abstract}
It is argued that electron can be treated as an anyon which carries a charge
(-e) and a magnetic flux $\pm \frac{\Phi _{0}}{2}$ in the presence and
absence of a uniform external magnetic field. This flux is shown to arise
due to the spin of the electron. The flux associated with the electron spin
is calculated using a semi-classical model which is based on the magnetic
top model. In accordance with spherical top model it is assumed that the
spin angular momentum of the electron is produced by the fictitious point
charge (-e) rotating in a circular orbit. It is shown that the flux through
the circular orbit is independent of the radius and $\frac{\Phi _{0}}{2}$
for a spin down electron and -$\frac{\Phi _{0}}{2}$ for a spin up one. Where 
$\Phi_{0}=\frac{hc}{e}$ is the flux quantum.
\end{abstract}


\section{Introduction}

In two dimensional space, quantum particle, can exhibit a continuous range
of statistics interpolating between bosons and fermions \cite
{Leinaas,Wil,Wilc}. Such exotic particles are called anyons and they are
believed to play a role in the Fractional Quantum Hall Effect and, perhaps,
even in high-temperature superconductors \cite{Wu}. Wilczek \cite{Wil,Wilc}
showed that, each anyon is taken to be a boson carrying a unit statistical
magnetic flux, $\Phi $, concentrated at its location. When one anyon circles
once around another in an anti-clockwise sense; the wavefunction picks up a
Aharanov-Bohm phase: $\exp \left[ -2\pi i\frac{2e\phi }{hc}\right] \equiv
\exp (2i\theta ).$ The cases $\theta =0,\frac{\pi }{2}$ and $\pi $
correspond to the particles being bosons, semions and fermions respectively 
\cite{Chitra}. Alternatively, we can take the anyons to be fermions carrying
a statistical flux 2$\theta $. Then $\theta =0,\pi /2$ and $\pi $ correspond
to fermions, semions and bosons. The purpose of this note is to present that
electron itself may be considered as an anyon. Namely it carries a charge
and a magnetic flux associated with its spin. We calculate the magnetic flux
associated with the electron's spin using by a semi-classical model which is
based on the magnetic top model \cite{Rosen,Schulman,Barut} that can be made
equivalent to a circular motion of a point charge in two dimensions. We show
that the magnetic flux associated with electron spin is $\frac{hc}{2e}$ for
spin down electron and -$\frac{hc}{2e}$ for spin up one.

\section{Formalism}

Generally the Lagrangian of a non-relativistic electron with mass m and
electric charge (-e) moving in a uniform magnetic field in z direction ($%
\overrightarrow{B}=B\widehat{k}$) is given by

\begin{equation}
L=\frac{1}{2}m\overrightarrow{v}^{2}-\frac{e}{c}\overrightarrow{v}.%
\overrightarrow{A}(r)
\end{equation}
where $\overrightarrow{r}=\overrightarrow{r}(x,y)$ is the position vector in
2D (two dimensions), $\overrightarrow{v}=\stackrel{\cdot }{\overrightarrow{r}%
}$ is the velocity vector and $\overrightarrow{A}$ is the vector potantial ($%
\overrightarrow{B}=\overrightarrow{\nabla }\times \overrightarrow{A}$). In
the symmetric gauge, the vector potantial can be written as

\begin{equation}
\overrightarrow{A}=\frac{1}{2}\overrightarrow{B}\times \overrightarrow{r}=%
\frac{B}{2}(-y\widehat{i}+x\widehat{j})
\end{equation}
where $\widehat{i}$ and $\widehat{j}$ are the unit vectors along the x and y
axis respectively.

To calculate the flux in terms of the radius and the magnetic field; we
write:

\begin{equation}
\Phi =\oint \frac{\overrightarrow{B}}{2}.(\overrightarrow{r}\times 
\overrightarrow{dr})=\oint \frac{\overrightarrow{B}}{2}.(\overrightarrow{r}%
\times \frac{\overrightarrow{dr}}{dt})dt
\end{equation}

If the electron is rotating in x-y plane in the counter clockwise direction
with the angular frequency $\omega _{c}=\frac{eB}{mc}$. The position vector
is

\begin{equation}
\overrightarrow{r}=r\cos \omega _{c}t\widehat{i}+r\sin \omega _{c}t\widehat{j%
}.
\end{equation}

Taking the time derivative of Eq.(4) we get the velocity vector:

\begin{equation}
\overrightarrow{v}=\frac{\overrightarrow{dr}}{dt}=-r\omega _{c}\sin \omega
_{c}t\widehat{i}+r\omega _{c}\cos \omega _{c}t\widehat{j}.
\end{equation}

Substitution of Eqs.(4),(5) and $\overrightarrow{B}=B\widehat{k}$ in Eq.(3)
we get

\begin{equation}
\Phi =\frac{B}{2}\omega _{c}r^{2}\int\limits_{0}^{T_{c}=\frac{2\pi }{\omega
_{c}}}dt=\pi r^{2}B=\pi (x^{2}+y^{2})B
\end{equation}
where $\omega _{c}=\frac{2\pi }{T_{c}}=\frac{eB}{mc}$ is the cyclotron
angular frequency. We note that the time integral in Eq.(6) has to be taken
over one cyclic period $T_{c}=\frac{2\pi }{\omega _{c}}$. This is the
crucial point of our calculation.

By using Eqs.(6) and (2) the vector potantial, $\overrightarrow{A}$ \ in the
symmetric gauge can be related to the magnetic flux $\Phi $ by

\begin{equation}
\overrightarrow{A}(r)=\frac{\Phi }{2\pi }(\frac{-y}{x^{2}+y^{2}}\widehat{i}+%
\frac{x}{x^{2}+y^{2}}\widehat{j}).
\end{equation}

The canonical momentum $\overrightarrow{p}$ can be derived from the
Lagrangian given in Eq.(1):

\begin{equation}
\overrightarrow{p}=\frac{\partial L}{\partial v}=m\overrightarrow{v}-\frac{e%
}{c}\overrightarrow{A}.
\end{equation}

Because of the \ rotationally invariance of the Lagrangian the canonical
angular momentum, J$_{c}$ is constant:

\begin{equation}
J_{c}=\overrightarrow{r}\times \overrightarrow{p}=\overrightarrow{r}\times (m%
\overrightarrow{v}-\frac{e}{c}\overrightarrow{A})=\overrightarrow{r}\times m%
\overrightarrow{v}-\frac{e}{c}\overrightarrow{r}\times \overrightarrow{A}=J-%
\frac{e\Phi }{2\pi c}
\end{equation}
where J is the gauge invariant kinetic angular momentum. This conserved
canonical angular momentum J$_{c}$ has a conventional spectrum \cite
{Lerda,Jackiw}: its eigenvalues are always integers in units of $\hbar $.
The difference between J$_{c}$ and J is due to presence of the magnetic
flux, and hence the magnetic field B. Both in the absence and in the
presence of the magnetic field B, the canonical angular momentum is always
represented by the quantum mechanical operator:

\begin{equation}
J_{c}=-i\hbar \frac{\partial }{\partial \varphi }
\end{equation}
where $\varphi $ is the polar angle on x-y plane.

The eigenvalues of J$_{c}$ is m$\hbar $ (m$\in Z$) when it acts on single
valued wavefunctions with angular dependence $\exp (im\varphi )$. Therefore
the kinetic angular momentum operator can be written as

\begin{equation}
J=J_{c}+\frac{e\Phi }{2\pi c}=-i\hbar \frac{\partial }{\partial \varphi }+%
\frac{e\Phi }{2\pi c}
\end{equation}
which when acting on single-valued wavefunctions with angular dependence $%
\exp (im\varphi )$ becomes:

\begin{equation}
J=\hbar (m+\frac{e\Phi }{hc})\text{ \ \ \ \ \ \ \ \ \ \ }(\text{m}\in Z)
\end{equation}
which states that the spectrum of J consists of integers shifted by $\frac{%
e\Phi }{hc}$ and is non-zero even for m=0. We believe that this non-zero
contributions to the kinetic angular momentum comes from spinning motion of
the electron. In the following part we calculate the magnetic flux
associated with the electron's spin using by a semi-classical model which is
based on the magnetic top model \cite{Rosen,Schulman,Barut} that can be made
equivalent to a circular current loop. Our calculations gives that magnetic
flux associated with electron spin is $\frac{hc}{2e}$ for a spin down
electron and -$\frac{hc}{2e}$ for spin up one.

As clearly stated by Barut et al \cite{Barut} the magnetic top is an
adequate model of quantum spin, because the magnetic moment performs a
simple precession around the magnetic field although the top itself performs
a complicated motion. it was also shown that by canonical and
Schr\"{o}dinger quantization, Pauli theory of spin was obtained. In the
magnetic top model the electron is assumed to be \ a small sphere with the
radius r$_{e}$ and the spin of the electron is assumed to be produced by the
electron's rotation about itself with an angular frequency $\omega _{s}$.
Our model \cite{Saglam} is based on the magnetic top model which can be made
equivalent to a circular current loop with the radius R in x-y plane. It
will be shown that as far as the flux is concerned the radius R of this loop
is a phenomenal concept and it gets eliminated in the end. In this model the
electron motion is considered in two parts namely an ''external'' motion
which can be interpreted as the motion of the center of mass (and hence the
central of charge) and an ''internal'' one whose average disappears in the
calssical limit. The latter is caused by the spin of the electron. The
important thing is that although the average of the internal motion
disappears the average of the flux associated with the internal motion does
not. In accordance with the magnetic top model, we assume that the spin
angular momentum of the electron is produced by the fictitious point charge
(-e) rotating in a circular orbit with radius R and with the same angular
frequency $\omega _{s}$. It can be shown that the present calculations can
be generalized to any kind of spherical charge distribution for electron.
The details of the charge distribution determines the relation between the
radius r$_{e}$ of the electron and the radius R of the current loop. Further
it is important to note that electron's spinning frequency $\omega _{s}$ is
very high compared to the cyclotron frequency $\omega _{c}$ (for B=5x10$^{3}$
Gauss, $\omega _{s}\approx 10^{15}\omega _{c}$ ). As was shown in Eq.(6) for
calculation of the magnetic flux the cyclotron period $T_{c}=\frac{2\pi }{%
\omega _{c}}$ is the important time interval. During the cyclotron period $%
T_{c}$, electron completes only one turn around its cyclotron orbit, but it
spins ($\frac{\omega _{s}}{\omega _{c}}>>1$) times about itself. Although
the radius of the electron ( and hence the radius of the loop) is very
small, because of \ the rapid spinning, the total flux during the cyclotron
period, $T_{c}$ will be comperable with the flux quantum, $\Phi =\frac{hc}{e}
$. We will see that the total flux associated with the spin is exactly $\pm 
\frac{\Phi _{0}}{2},$ where (+) sign stands for spin down electron and (-)
sign for spin up one.

\bigskip In the above described model we define the vector going from origin
to the fictitious point charge (-e) as

\begin{equation}
\overrightarrow{r}^{^{\prime }}=\overrightarrow{r}+\overrightarrow{R}
\end{equation}
where $\stackrel{\rightarrow }{r}$ is the vector going from origin to the
centre of mass of the electron and $\stackrel{\rightarrow }{R}$ is the
vector going from the centre of mass to this fictitious point charge (-e)
rotating in a circular orbit with a radius $\stackrel{\rightarrow }{R}$ and
an angular frequency $\omega _{s}$. So the vectors $\stackrel{\rightarrow }{R%
}(\uparrow )$ and $\stackrel{\rightarrow }{R}(\downarrow )$ for spin up and
down electrons read:

\begin{equation}
\stackrel{}{\stackrel{\rightarrow }{R}(\uparrow )=R\cos \omega _{s}t%
\stackrel{\wedge }{i}-}R\sin \omega _{s}t\stackrel{\wedge }{j}
\end{equation}

\begin{equation}
\stackrel{\rightarrow }{R}(\downarrow )=R\cos \omega _{s}t\stackrel{\wedge }{%
i}+R\sin \omega _{s}t\stackrel{\wedge }{j}.
\end{equation}

From Eqs.(4) and (14) the vector $\stackrel{\rightarrow }{r^{\prime }}$ for
spin up electron reads,

\begin{eqnarray}
\stackrel{}{\stackrel{\rightarrow }{r^{\prime }}(\uparrow )} &=&\stackrel{%
\rightarrow }{r}+\stackrel{\rightarrow }{R(\uparrow )}=[r\cos \omega
_{c}t+R\cos \omega _{s}t]\stackrel{\wedge }{i} \\
&&+[r\sin \omega _{c}t-R\sin \omega _{s}t]\stackrel{\wedge }{j}  \nonumber
\end{eqnarray}

The time derivation of Eq.(16) gives

\begin{eqnarray}
\frac{\stackrel{\rightarrow }{dr^{\prime }}(\uparrow )}{dt} &=&[-r\omega
_{c}\sin \omega _{c}t-R\omega _{s}\sin \omega _{s}t]\stackrel{\wedge }{i} \\
&&+[r\omega _{c}\cos \omega _{c}t-R\omega _{s}\cos \omega _{s}t]\stackrel{%
\wedge }{j}  \nonumber
\end{eqnarray}

Analogously to Eq.(3) the total flux $\Phi ^{\prime }(\uparrow )$ for spin
up electron is

\begin{equation}
\Phi ^{\prime }(\uparrow )=\oint \frac{\stackrel{\rightarrow }{B}}{2}.[%
\stackrel{}{\stackrel{\rightarrow }{r^{\prime }}(\uparrow )}\times \stackrel{%
}{\stackrel{\rightarrow }{dr^{\prime }}(\uparrow )}]=\frac{\stackrel{%
\rightarrow }{B}}{2}.\oint\limits_{0}^{T_{c}}\stackrel{\rightarrow }{%
r^{\prime }}(\uparrow )\times \frac{\stackrel{\rightarrow }{dr^{\prime }}}{dt%
}dt
\end{equation}

Substitution of Eqs.(16) and (17) into Eq.(18) gives:

\begin{equation}
\Phi ^{\prime }(\uparrow )=\frac{\stackrel{}{B}}{2}\oint\limits_{0}^{T_{c}}(%
\omega _{c}r^{2}-\omega _{s}R^{2}+\text{cross terms})dt
\end{equation}
where we used the identity: $\sin ^{2}x+\cos ^{2}x=1.$ Here the cross terms
contain the product of different angular frequencies and phases, so the
integral of these terms vanishes and Eq.(19) reduces to

\begin{equation}
\Phi ^{\prime }(\uparrow )=\pi r^{2}B-\frac{\omega _{s}}{\omega _{c}}\pi
R^{2}B
\end{equation}

With a similar procedure the flux for spin down electron becomes

\begin{equation}
\Phi ^{\prime }(\downarrow )=\pi r^{2}B+\frac{\omega _{s}}{\omega _{c}}\pi
R^{2}B
\end{equation}
In Eqs.(20) and (21) the first terms are the fluxes without the electron
spin [Eq.(6)]. While the second terms are the spin contributions.

Now we want to calculate the second terms in terms of the flux quantum $\Phi
_{0}$: The spin magnetic moment $\stackrel{\rightarrow }{\mu }$ of a free
electron is given by 
\begin{equation}
\stackrel{\rightarrow }{\mu }=-g\stackrel{}{\mu _{B}}\stackrel{\rightarrow }{%
S}
\end{equation}
where $\hbar \stackrel{\rightarrow }{S}$ is the spin angular momentum of the
electron. When we introduce the magnetic field $\stackrel{\rightarrow }{B}=B%
\stackrel{\wedge }{k},$ the z component of the magnetic moment becomes:

\begin{equation}
\mu _{z}=\pm \mu _{B}=\pm \frac{e\hbar }{2mc}.
\end{equation}
where we put g=2 for a free electron.

As we stated earlier we assume that the spin angular momentum of the
electron is produced by the fictitious point charge (-e) rotating in a
circular orbit with the angular frequency $\omega _{s}$ and the radius R in
x-y plane. In this case the z-component of this magnetic moment for spin
down electron will be

\begin{equation}
\mu _{z}=-\frac{IA}{c}=-\frac{e\omega _{s}R^{2}}{2c}
\end{equation}
where we put I=$\frac{e\omega _{s}}{2\pi }$ and A=$\pi R^{2}.$

If we compare Eqs.(23) and (24) we find

\begin{equation}
\omega _{s}=\frac{\hbar }{mR^{2}}
\end{equation}
which relates the spinning angular frequency $\omega _{s}$ to the radius R.

Substitution of Eq.(25) and $\omega _{c}=\frac{eB}{mc}$ in the second terms
of Eqs.(20) and (21) we find

\begin{equation}
\pm \frac{\omega _{s}}{\omega _{c}}\pi R^{2}B=\pm \frac{hc}{2e}=\pm \frac{%
\Phi _{0}}{2}
\end{equation}

Eq.(26) states that the flux associated with the electron spin is
independent of the radius of circular loop ( $\varpropto $ the radius of the
electron) and the mass of \ the electron. The (+) sign stands for spin down
electron and (-) sign for spin up one.

Next we want to calculate the first terms \ of Eq.(20) and (21) in terms of
the flux quantum. To do so we look at the Schr\"{o}dinger equation for an
electron in the presence of the uniform magnetic field $\overrightarrow{B}=B%
\widehat{k}$. Let the center of the electron's orbit be the center of a
cylindirical coordinate system $(\rho ,\phi ,z)$ in which the z-axis is
directed along the magnetic field $\stackrel{\rightarrow }{B}$. If we take
the vector potantial described in Eq.(2) namely$\ \stackrel{\rightarrow }{A}=%
\frac{1}{2}\stackrel{\rightarrow }{B}\times \stackrel{\rightarrow }{r}$ ,
ie. $A_{\phi }=\frac{1}{2}B\rho $,\ $A_{\rho }=A_{z}=0.$ Then the
Schr\"{o}dinger equation in this gauge becomes:

\begin{equation}
-\frac{\hbar ^{2}}{2m}\left[ \frac{1}{\rho }\frac{\partial }{\partial \rho }%
\left\{ \rho \frac{\partial \Psi }{\partial \rho }\right\} +\frac{\partial
^{2}\Psi }{\partial z^{2}}+\frac{1}{\rho ^{2}}\frac{\partial ^{2}\Psi }{%
\partial \phi ^{2}}\right] -\frac{ie\hbar B}{2mc}\frac{\partial \Psi }{%
\partial \phi }+\frac{e^{2}B^{2}\rho ^{2}}{8mc^{2}}\Psi =E\Psi
\end{equation}
Because of the cylindirical symmetry all terms containing $\frac{\partial }{%
\partial \phi }$ vanish and since the electron is assumed to be moving in
x-y plane $\frac{\partial ^{2}\Psi }{\partial z^{2}}=0$. So Eq.(27) is
reduced to a simpler equation of polar coordinates:

\begin{equation}
-\frac{\hbar ^{2}}{2m}\left[ \frac{1}{r}\frac{\partial }{\partial r}\left( r%
\frac{\partial \Psi }{\partial r}\right) \right] +\frac{\hbar ^{2}r^{2}}{%
8m\lambda ^{4}}\Psi =E\Psi
\end{equation}
where

\begin{equation}
\rho \equiv r=\sqrt{x^{2}+y^{2}}.
\end{equation}
The solutions of Eq.(28) has the form:

\begin{equation}
\Psi _{n}(r)\propto H_{n}(r)e^{-\frac{r^{2}}{4\lambda ^{2}}}
\end{equation}
where $H_{n}(r)$ is the n$^{th}$ Hermite polynomial. The eigenvalues
corresponding to Eq.(30) are

\begin{equation}
E_{n}=(n+\frac{1}{2})\hbar \omega _{c}=(n+\frac{1}{2})\frac{\hbar eB}{mc}%
\text{ \ \ \ \ \ \ \ \ \ (n=0,1,2,...)}
\end{equation}
which are the well-known Landau levels.

\bigskip The probability density P(r) in polar coordinates is given by:

\begin{equation}
P(r)=\left| \Psi _{n}(r)\right| ^{2}2\pi r
\end{equation}
which has its maxima at

\begin{equation}
r_{n}=\sqrt{2n+1}\lambda
\end{equation}
From Eq.(33) the magnetic flux corresponding to these orbitals will be:

\begin{equation}
\Phi _{n}=\pi r_{n}^{2}B=(n+\frac{1}{2})\frac{hc}{e}=(n+\frac{1}{2})\Phi
_{0}.
\end{equation}

If we compare Eqs.(31) and (34) we find that without spin, magnetic energy
and the flux are proportional to each other:

\begin{equation}
\Phi _{n}=\frac{2\pi c}{e\omega _{c}}E_{n}.
\end{equation}

It is tempting to think that this proportionality is valid even in the
presence of spin (ie. in the presence of Zeeman term in the energy). So in
the presence of spin the additional Zeeman term ($\pm \frac{1}{2}g\mu _{B}B$%
) in the energy should produce an additional flux term $\frac{2\pi c}{%
e\omega _{c}}(\pm \frac{1}{2}g\mu _{B}B)=\pm \frac{\Phi _{0}}{2}$ which is
exactly what we derived in Eq.(26). Another test of our model is that the
definition of the spin of an anyon: Generally the spin of an anyon is
defined \cite{Wil,Wilc,Lerda} by

\begin{equation}
s=\frac{J(m=0)}{\hbar }=\frac{e\Phi }{hc}.
\end{equation}

If we substitute the value of $\Phi $ from Eq.(26), we find:

\begin{equation}
s=\frac{e\Phi }{hc}=\pm \frac{1}{2}
\end{equation}
which is exactly equal to the electron spin.

So far we have considered only the flux associated with the external
magnetic field. However even in the absence of the external field, because
of its spin, electron will create and carry a magnetic flux with itself. In
accordance with the magnetic top model, we can assume that the spin angular
momentum of the electron is produced by the fictitious point charge (-e)
rotating in a circular orbit with radius R and angular frequency $\omega
_{s} $ \cite{Saglam}. If we take the spin vector in z direction then the
magnetic field lines associated with the spin magnetic moment, $%
\overrightarrow{\mu }_{e}$ will be as in Fig.1a.

\epsfig{file=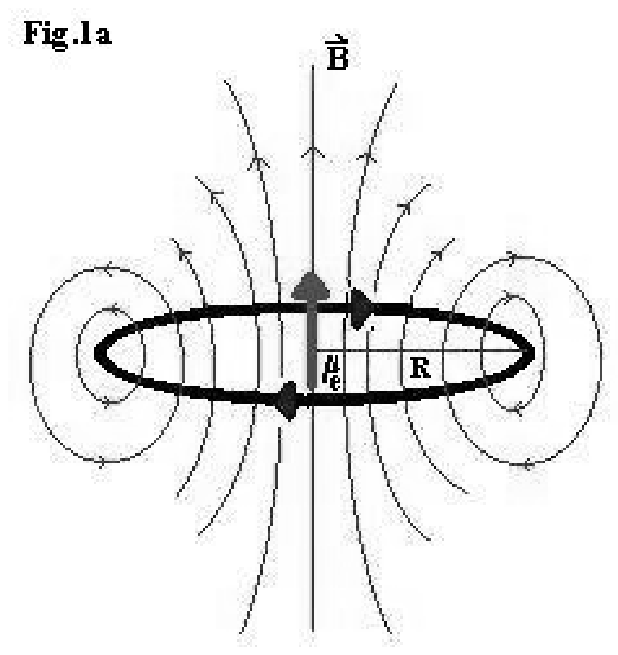} \epsfig{file=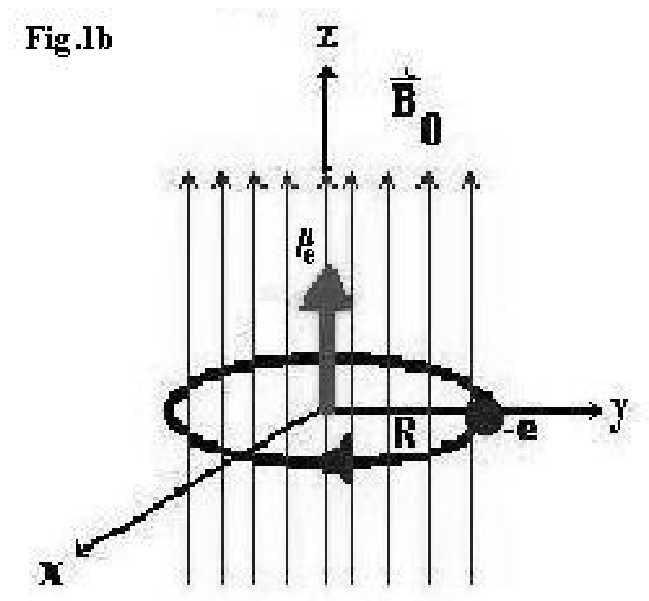}

Outside the current loop the magnetic vector potantial $\overrightarrow{A}%
_{e}$ of the spin magnetic moment $\overrightarrow{\mu }_{e}$ is given by

\begin{equation}
\stackrel{\rightarrow }{A_{e}}=\frac{\stackrel{\rightarrow }{\mu _{e}}\times 
\stackrel{\rightarrow }{r}}{r^{3}}\text{ \ \ \ \ \ \ \ \ }r\geq R
\end{equation}

The magnetic field $\overrightarrow{B}$ of the electron is calculated
through the relation:

\begin{equation}
\stackrel{\rightarrow }{B}=\stackrel{\rightarrow }{\nabla }\times \stackrel{%
\rightarrow }{A}_{e}
\end{equation}

In the x-y plane the x and y components of $\stackrel{\rightarrow }{B}$
vanish and the z component takes the form

\begin{equation}
B_{z}=-\frac{\mu _{e}}{r^{3}}\text{ \ \ \ \ \ \ \ \ \ \ \ \ }r\geq R
\end{equation}
where $r=\sqrt{x^{2}+y^{2}.}$

We want to calculate the fluxes through the x-y plane outside and inside the
current loop with radius R: The related fluxes will be defined as follows:

$\Phi _{out}$ (R)= The total flux through x-y plane outside the loop (r $%
\geq $ R)

$\Phi _{in}$(R) = The total flux through x-y plane inside the loop (r $<$ R)

The condition of no magnetic monopoles ($\stackrel{\rightarrow }{\nabla }.%
\stackrel{\rightarrow }{B}=0$) requires that the number of magnetic field
lines inside the loop must be equal to the number of magnetic field lines
outside the loop. Therefore

\begin{equation}
\Phi _{in}(R)=-\Phi _{out}(R)
\end{equation}

$\Phi _{out}(R)$ can be calculated easily from eq.(40):

\begin{equation}
\Phi _{out}(R)=\int\limits_{R}^{\infty }\stackrel{\rightarrow }{B}.\stackrel{%
\rightarrow }{da}=\int\limits_{R}^{\infty }\frac{\mu _{e}}{r^{3}}2\pi rdr=%
\frac{2\pi \mu _{e}}{R}
\end{equation}

where $\overrightarrow{da}=-2\pi rdr\widehat{k}$ is used to be consistent
with the previous area vectors.From Eq.(41) $\Phi _{in}$(R) will be

\begin{equation}
\Phi _{in}(R)=-\frac{2\pi \mu _{e}}{R}
\end{equation}

In the present current loop model it can be shown that inside the loop the
z-component of the magnetic field is the same at every point and is equal to
the magnetic field at the center, ie. for $r<R$.

\begin{equation}
B_{z}=B_{0}=\frac{2\pi I}{cR}=\frac{2\mu _{e}}{R^{3}}
\end{equation}
where we used $I=\frac{e\omega _{s}}{2\pi }$ and $\mu _{e}=\frac{IA}{c}=%
\frac{I\pi R^{2}}{c}.$ This can be proved just by calculating the flux of
uniform field $\overrightarrow{B}_{0}$ through the loop: 
\begin{equation}
\Phi _{in}(R)=\oint \stackrel{\rightarrow }{B_{0}}.\frac{\overrightarrow{R}%
(\uparrow )\times \text{ }\overrightarrow{dR}(\uparrow )}{2}%
=\oint\limits_{0}^{T_{s}}\frac{\stackrel{\rightarrow }{B_{0}}}{2}.(%
\overrightarrow{R}(\uparrow )\times \frac{\overrightarrow{dR}(\uparrow )}{dt}%
)dt
\end{equation}

Substitution of Eq.(14) and its derivative in Eq.(45) gives

\begin{equation}
\Phi _{in}(R)=-\frac{B_{0}}{2}\omega _{s}R^{2}T_{s}=-\frac{2\pi \mu _{e}}{R}
\end{equation}
where we used Eq.(44) and put $T_{s}=\frac{2\pi }{\omega _{s}}.$ Therefore
as far as the magnetic flux is concerned the field lines inside the loop can
be replaced by the field lines of the uniform magnetic field $%
\overrightarrow{B}_{0}=\frac{2\mu _{e}}{R^{3}}\widehat{k}$. So in the
equvalent picture of $\Phi _{in}(R)$ we have a uniform magnetic $B_{0}$ and
a fictitious point charge (-e) rotating in the clockwise sense as in Fig.1b.

From Eq.(45) we can state that the flux $\Phi _{in}(R)$ is the one that
corresponds to spinning period $T_{s}=\frac{2\pi }{\omega _{s}}.$ On the
other hand the uniform magnetic field $B_{0}$ inside the loop defines
another period $T_{c}$ which is the cyclotron period:

\begin{equation}
T_{c}=\frac{2\pi }{\omega _{c}}=\frac{2\pi mc}{eB_{0}}=\frac{\pi mcR^{3}}{%
e\mu _{e}}
\end{equation}

We have seen that for flux quantization the cyclotron period $T_{c}$ of the
magnetic field is the important time interval. Therefore the total flux
corresponding to $T_{c}$ is

\begin{equation}
\Phi (\uparrow )=\frac{T_{c}}{T_{s}}\Phi _{in}(R)=-\frac{T_{c}}{T_{s}}\frac{%
2\pi \mu _{e}}{R}
\end{equation}

Substitution of Eq.(25) and (47) gives

\begin{equation}
\Phi (\uparrow )=-\frac{hc}{2e}=-\frac{\Phi _{0}}{2}.
\end{equation}

With a similar prosedure the flux associated with the spin down electron
will be

\begin{equation}
\Phi (\downarrow )=\frac{hc}{2e}=\frac{\Phi _{0}}{2}.
\end{equation}

Therefore even in the absence of the external magnetic field the flux
associated with electron spin will be $\pm \frac{\Phi _{0}}{2}$.

\section{Results and Discussion}

We have calculated the magnetic flux associated with the electron spin to
show that electron itself can be treated as an anyon. Our model is based on
the magnetic top model which can be made equivalent to a circular current
loop with radius R. As far as the flux is concerned the radius R of this
loop is a phenomenal concept and is eliminated in the end. We distinguish
the spin angular frequency, $\omega _{s}$ from the cyclotron angular
frequency $\omega _{c}=\frac{eB}{mc}(\omega _{s}>>\omega _{c}).$ It is shown
that the calculation of the flux is led to a time integral. The crucial
point is that, the limits of time is taken from zero to cyclotron period, $%
T_{c}$. During the cyclotron period $T_{c}$, electron completes one turn
around the cyclotron orbit, but it spins ($\omega _{s}/\omega _{c}$) times
about itself. Although the radius of the electron is very small, because of
the rapid spinning, the total flux associated with the electron spin is
comparable with the flux quantum. As far as the flux is concerned the mass
of the electron is a parameter which is eliminated in the end. Therefore
whether we take the relativistic mass or not the result does not change (to
go from Eq.(22) to (26) we do not put any restriction on the electronic
mass). Namely the flux is $\pm \frac{\Phi _{0}}{2}=\pm \frac{hc}{2e}$. It
can be shown that the present calculations can be generalize to any
spherical charge distribution for electron. The details of the charge
distribution determines the relation between the radius of the electron r$%
_{e}$ and the radius R of the equivalent current loop. For a uniform charge
distribution the magnetic moment of the spinning electron is found to be $%
\mu =\frac{e\omega _{s}r_{e}^{2}}{5c}$ if we compare this with Eq.(24), the
relation between R and r$_{e}$ becomes: $R=\sqrt{\frac{2}{5}}r_{e}.$ We
believe that the present study will bring a new insight to understand the
area of physics which concerns the magnetic flux. A more complete report
will be given in future.

\section{Acknowledgements}

The author would like to thank to Prof. A. Vercin for valuable discussions.


\begin{references}
\bibitem{Leinaas}  Leinaas J. and Myrheim I., Nuova Cimento, {\bf B37}, 1,
(1977).

\bibitem{Wil}  Wilczek, F., Phys. Rev. Lett., {\bf 48}, 1144, (1982).

\bibitem{Wilc}  Wilczek, F., Phys. Rev. Lett., {\bf 49}, 957, (1982).

\bibitem{Wu}  Wu, Y.-S., Phys. Lett., {\bf 53}, 111, (1984).

\bibitem{Chitra}  Chitra, R., Modern Phys. Lett. {\bf A7}, No:10, 855,
(1992).

\bibitem{Rosen}  Rosen, N., Phys. Rev. {\bf 82}, 621, (1951).

\bibitem{Schulman}  Schulman, L., Phys. Rev. {\bf 176}, 1558, (1968).

\bibitem{Barut}  Barut, A.O.,Bozic, M. and Maric, Z., Annals of Phys. {\bf %
214}, 53, (1992).

\bibitem{Lerda}  Lerda, A., Anyons. Springer-Verlag Heidelberg. (1992).

\bibitem{Jackiw}  Jackiw, R. and Redlich, A. N., Phys. Rev. Lett., {\bf 50},
555, (1983)

\bibitem{Saglam}  Saglam, M. and Boyacioglu, Proc. 9th Internat. Conf. on
Hopping and Related Phenomena, Shefayim (Israel), Sept. 3-6, 2001, to be
published in Phys. Stat. Sol. (b) (2002).
\end{references}
\end{document}